\documentstyle[epsfig]{nature-camilo}
\newcommand {\ga}{\ {\raise-.5ex\hbox{$\buildrel>\over\sim$}}\ }
\newcommand {\la}{\ {\raise-.5ex\hbox{$\buildrel<\over\sim$}}\ }
\newcommand{\columbia}{\footnotesize Columbia Astrophysics Laboratory,
Columbia University, 550 West 120th Street, New York, New York 10027, USA.}
\newcommand{\nrao}{\footnotesize National Radio Astronomy Observatory,
520 Edgemont Road, Charlottesville, Virginia 22903, USA.}
\newcommand{\parkes}{\footnotesize Australia Telescope National Facility,
CSIRO, Parkes Observatory, PO Box 276, Parkes, New South Wales 2870, Australia.}
\newcommand{\xte}{XTE~J1810--197}

\title{Transient pulsed radio emission from a magnetar}

\author{Fernando Camilo\affiliation{\columbia}, Scott M.\
Ransom\affiliation{\nrao}, Jules P.\ Halpern~$^{1}$, John
Reynolds\affiliation{\parkes}, David J.\ Helfand~$^{1}$, Neil
Zimmerman~$^{1}$ \& John Sarkissian~$^{3}$\\}

\headertitle{Transient pulsed radio emission from a magnetar}
\mainauthor{Camilo et~al.}

\summary{\noindent Anomalous X-ray pulsars (AXPs) are slowly rotating
  neutron stars with very bright and highly variable X-ray emission
  that are believed to be powered by ultra-strong magnetic fields
  of $>10^{14}$\,G, according to the `magnetar' model\cite{dt92a}.
  The radio pulsations that have been observed from more than 1,700
  neutron stars with weaker magnetic fields have never been detected
  from any of the dozen known magnetars.  The X-ray pulsar \xte\ was
  revealed (in 2003) as the first AXP with transient emission when its
  luminosity increased 100-fold from the quiescent level\cite{ims+04};
  a coincident radio source of unknown origin was detected one year
  later\cite{hgb+05}.  Here we show that \xte\ emits bright, narrow,
  highly linearly polarized radio pulses, observed at every rotation,
  thereby establishing that magnetars can be radio pulsars.  There is
  no evidence of radio emission before the 2003 X-ray outburst (unlike
  ordinary pulsars, which emit radio pulses all the time), and the
  flux varies from day to day.  The flux at all radio frequencies is
  approximately equal --- and at $>20$\,GHz \xte\ is currently the
  brightest neutron star known.  These observations link magnetars to
  ordinary radio pulsars, rule out alternative accretion models for AXPs,
  and provide a new window into the coronae of magnetars.  }

\dates{12 May}{14 June 2006}

\begin{document}
\maketitle

We observed the position of \xte\ (ref.~4\nocite{irm+04}) on 17~March
2006 for 4.2 hours with the Parkes telescope at a central frequency
$\nu = 1.4$\,GHz, using parameters identical to those used in the Parkes
multibeam pulsar survey\cite{mlc+01}.  Pulsations with period $P=5.54$\,s
were easily detected, with period-averaged flux density $S_{1.4} = 6$\,mJy
and a narrow average profile with full-width at half-maximum of 0.15\,s
(Fig.~\ref{fig:profs}).  We detected individual pulses from virtually
every rotation of the neutron star (see Fig.~\ref{fig:sp}).  These are
composed of $\la 10$-ms--wide sub-pulses with peak flux densities up to
$\ga 10$\,Jy and follow a differential flux distribution approximated
by $d\,\log N = - d\,\log S$, with no giant pulses like those observed
from the Crab pulsar\cite{cbh+04}.

A timing model accounting for every turn of the neutron star during the
period 17 March--7 May yields barycentric $P = 5.54024870\,\mbox{s} \pm
20$\,ns on MJD 53855.0 and $\dot P = (1.016\pm 0.001)\times10^{-11}$,
with root-mean-square residual $\sigma = 5$\,ms.  We use this to
set constraints on any putative companion to the AXP by requiring the
light-travel-time across the projected orbital semi-major axis to be less
than $\sigma$.  From Kepler's third law, with assumed neutron star mass
$1.4$\,M$_\odot$, the upper limits on the minimum companion mass lie in
the range $\sim 0.003$--$0.03$\,M$_\odot$ for orbital periods in the
range 2\,h--5\,min, effectively ruling out the existence of any Roche
lobe-filling star orbiting this AXP.  The delay in pulse arrival times
measured between 2.9 and 0.7\,GHz implies an integrated column density of
free electrons between the Earth and \xte\ of $178\pm5$\,cm$^{-3}$\,pc.
Together with a model for the Galactic distribution of free
electrons\cite{cl02}, the distance to \xte\ is $D\approx 3.3$\,kpc
(here we use $\approx$ to indicate a quantity known to within about a
factor of two or better), consistent with X-ray- and optically-derived
estimates of 2.5--5\,kpc (refs~8--10)\nocite{gh05,ghbb04,dvk06}.

In the original detection of radio emission from \xte\ with the
Very Large Array (VLA) in January 2004, $S_{1.4} = 4.5$\,mJy
(ref.~3)\nocite{hgb+05}.  In February 2006 at the VLA we measured
$S_{1.4} = 12.9$\,mJy, a considerable increase in flux over a period of
two years.  However, further observations (see Table~\ref{tab:fluxes})
reveal large fluctuations: at Parkes, $S_{1.4}$ can vary by factors of
about two from one day to the next, although no significant variations
are detected within observing sessions lasting up to four hours.
These flux variations are completely inconsistent with expectations from
interstellar scintillation\cite{ric90,nar92a} and are thus, remarkably
for a pulsar, intrinsic to \xte.  At the VLA, $S_{8.4}$ shows variations
by similar factors within 20\,min.  This more rapid variation at high
frequencies is confirmed in pulsed observations with the Green Bank
Telescope (GBT).  However, at $\nu \ga 9$\,GHz diffractive interstellar
scintillation\cite{ric90,nar92a} plays a significant role\cite{cl02}.
At present we cannot rule out a non-pulsed component.  The maximum
size of any extended emission is $5\times10^{15}(D/3\,\mbox{kpc})$\,cm,
based on the 0.25-arcsec resolution of our VLA image at 8.4\,GHz, from
which we measure a position for \xte\ of right ascension $= 18\,{\rm
h}\,09\,{\rm min}\,51.087\,{\rm s} \pm 0.001\,{\rm s}$ and declination $=
-19^\circ\,43'\,51.93'' \pm 0.02''$, in the J2000.0 equinox.

The radio spectrum of \xte\ is quite flat over a factor of 60 in
frequency.  We have made independent simultaneous measurements of flux
at multiple frequencies (see Table~\ref{tab:fluxes}), all of which
are consistent with spectral index $\alpha \ga -0.5$, where $S_{\nu}
\propto \nu^{\alpha}$.  Ordinary pulsars have mean $\alpha = -1.6$, and
fewer than 10\% have $\alpha>-0.5$ (ref.~13)\nocite{lylg95}.  With an
unremarkable average flux at $\nu \sim 1$\,GHz (here we use $\sim$
to indicate a quantity known to within about an order of magnitude), by
$\nu \ga 20$\,GHz, \xte\ is brighter than every other known neutron star.
Its assumed isotropic radio luminosity up to 42\,GHz is about $2\times
10^{30}$\,erg\,s$^{-1}$, compared to the spin-down luminosity of about
$2.4\times 10^{33}$\,erg\,s$^{-1}$.

Radiation from \xte\ is highly polarized: $89\pm5\%$ of the
total-intensity 8.4-GHz flux measured at the VLA is linearly polarized.
Also, a 1.4-GHz Parkes observation shows linear polarization that tracks
the total-intensity pulse profile at a level of $65\%$.

\xte\ is located within the boundaries of the Parkes multibeam pulsar
survey area\cite{mlc+01}.  We have analysed archival raw survey data
from the two telescope pointings nearest to the pulsar position.
No pulsar signal was detected in average-pulse analyses, corresponding
to $S_{1.4} \la 0.2$\,mJy from these observations made in 1997 and 1998
(see Table~\ref{tab:fluxes}).  This limit is $<10\%$ of the flux from
the pulsar since 2004, suggesting that the extraordinary radio emission
from \xte\ turned on as a result of the events accompanying the X-ray
outburst observed in early 2003 (ref.~2)\nocite{ims+04}.

However, about 10\% of ordinary pulsars have specific radio luminosity
$L_{1.4} \equiv S_{1.4}D^2 \la 2$\,mJy\,kpc$^2$, the approximate limit
for \xte\ before the outburst.  For example, the pulsar PSR~J1718--3718,
with $L_{1.4} \approx 4$\,mJy\,kpc$^2$ and $P=3.3$\,s, has inferred
surface dipole magnetic field strength $B = 7\times10^{13}$\,G, higher
than that of one magnetar (for \xte, $B \approx 2.4\times10^{14}$\,G).
This source has X-ray properties apparently similar to those of \xte\
in quiescence\cite{km05}.  This raises the prospect that \xte\ could have
generated familiar radio pulsar emission before 2003, marking a plausible
direct link between ordinary pulsars and magnetars.  In addition, a
recently discovered class of radio-bursting neutron stars\cite{mll+06}
includes at least one object whose field strength is comparable to that
of some magnetars; it too has X-ray properties resembling those of \xte\
in quiescence\cite{rbg+06}.  We therefore also searched the archival
radio data for bright individual pulses, but found none.

Observable magnetospheric radio emission requires a coherent mechanism,
inferred to be curvature radiation whose characteristic frequency $3
\gamma^3 c/4 \pi r$ falls in the gigahertz range for secondary particle
Lorentz factors $\gamma \sim 10^3$ and radii of curvature $r$ of the
order of the light-cylinder radius $r_{\rm lc} = c P/2 \pi$ or smaller.
In ordinary pulsars, the required electron--positron pairs can be produced
only on the open field-line bundle at high altitude above the surface
where electric potentials of $\sim 10^{12}$\,V accelerate particles that
emit gamma-rays with energies exceeding the pair-creation threshold.
We cannot exclude that the handful of known magnetars employ this
mechanism, because their long periods imply small active polar caps
and narrow beams that may miss the observer for random orientations.
Additionally, magnetars have mechanisms and locations for pair creation
that are not available to ordinary pulsars\cite{tb05,bt06,zha01}.
Magnetar activity is generated by the sudden or continual twisting of the
external field lines into a non-potential configuration that maintains,
by induction, strong long-lived currents flowing along them\cite{tlk02}.
With required charge densities greatly exceeding the co-rotation
value\cite{gj69} for a dipole field, an electric potential of $\sim
10^9$\,V is established and self-regulated by pair creation.  In these
conditions, electrons and positrons are accelerated to $\gamma \sim 10^3$,
and pairs can be created either via resonant cyclotron scattering of
thermal X-rays in the strong magnetic field or thermally in the heated
atmosphere at the footpoints of the twisted field lines.  This affords
the possibility that magnetar radio emission is generated on closed
field lines and beamed into a wide range of angles.  Because most of the
energy of the twisted field is contained within a few stellar radii,
the frequencies of coherent emission from magnetars could be greater
than those of ordinary pulsars, possibly accounting for the flat radio
spectrum of \xte.  Also, it has been proposed that the observed optical
and infrared emission from magnetars is coherent emission from plasma
instabilities above the plasma frequency\cite{egl02,eltw03}, and even
that radio or submillimetre emission could be so generated\cite{lyu02}.
The extrapolated radio spectrum of \xte\ exceeds its observed infrared
fluxes\cite{rti+04}, so that the radio and infrared emission may or may
not have the same origin.

Radio emission as currently observed from XTE J1810--197 was evidently
not present during the historical quiescent X-ray state that persisted
for at least 24 years before the outburst\cite{hg05}.  Following this,
the onset of radio emission could have been delayed until the plasma
density declined to a value much less than is generally found in
persistent magnetars.  The hard and soft X-ray fluxes are now decaying
exponentially with timescales of $\approx 300$ and $\approx 900$ days,
respectively\cite{gh05}, while strong radio emission is still observed
more than three years after the X-ray turn-on.  We infer that radio
emission will cease after the transient X-ray components (and implied
magnetospheric currents) have subsided, but it is difficult to know
whether this transition is imminent, or will take many years.

\medskip
\noindent {\small {\bf Acknowledgements} \,\ We thank J.\ Cohen and B.\
Mason for giving us some of their observing time, M.\ Kramer and M.\
McLaughlin for providing us with Parkes multibeam survey archival data,
and D.\ Backer, D.\ Kaplan, and B.\ Jacoby for their contributions to
developing pulsar observing equipment at GBT.  The Parkes Observatory
is part of the Australia Telescope, which is funded by the Commonwealth
of Australia for operation as a National Facility managed by CSIRO.
The National Radio Astronomy Observatory is a facility of the US
National Science Foundation (NSF), operated under cooperative agreement
by Associated Universities, Inc.  F.C. acknowledges support from NSF
and NASA.

\medskip
\noindent {\small {\bf Author Information} \,\  Reprints and permissions
information is available at npg.nature.com/reprintsandpermissions.
The authors declare no competing financial interests. Correspondence
and requests for materials should be addressed to
F.C. (fernando@astro.columbia.edu).}

\clearpage

\begin{table*}[t]
\begin{center}
\begin{tabular}{lllc}
\\
\hline
Instrument  & Frequency, $\nu$ (GHz) & Date (MJD) &Flux density, $S_\nu$ (mJy)\\
\hline
Parkes      &  1.4                   & 50747      & $\la 0.25$                \\
Parkes      &  1.4                   & 51152      & $\la 0.15$                \\
VLA         &  1.4                   & 53019      &  $4.5\pm0.5$              \\
VLA         &  1.4                   & 53794      & $12.9\pm0.2$              \\
Parkes      &  1.4                   & 53811      &   6.0                     \\
Parkes      &  1.4                   & 53850.9$^a$&   6.8                     \\
Parkes      &  1.4                   & 53851.9    &  13.6                     \\
Parkes      &  1.4                   & 53852.9    &  13.2                     \\
Parkes      &  1.4                   & 53855.9$^b$&   8.7                     \\
GBT (BCPM)  &  1.4                   & 53862.4    &   5.1                     \\
Parkes      &  1.4                   & 53862.9    &   7.8                     \\
VLA         &  1.4                   & 53872      &  $3.3\pm0.2$              \\
VLA         &  1.4                   & 53874      &  $6.6\pm0.5$              \\
VLA         &  1.4                   & 53875      &  $4.5\pm0.5$              \\
VLA         &  1.4                   & 53877      &  $8.6\pm0.7$              \\
VLA         &  1.4                   & 53879      &  $5.8\pm0.6$              \\
GBT (BCPM)  &  1.9                   & 53857.3$^c$&   6.3                     \\
GBT (BCPM)  &  1.9                   & 53865      &   6.0                     \\
Parkes      &  0.69                  & 53850.9$^a$&   7.4                     \\
GBT (BCPM)  &  0.82                  & 53861.2    &   5.5                     \\
Parkes      &  2.9                   & 53850.9$^a$&   7.4                     \\
Parkes      &  6.4                   & 53855.9$^b$&   3.9                     \\
VLA         &  8.4                   & 53828      &  $9.0\pm0.1$              \\
GBT (BCPM)  &  9.0                   & 53857.4$^c$&   3.0                     \\
GBT (BCPM)  & 14                     & 53857.4$^c$&   2.1                     \\
GBT (Spigot)& 19                     & 53857.5$^c$&   1.9                     \\
GBT (Spigot)& 42                     & 53858.4    &   5.9                     \\
\hline
\end{tabular}
\end{center}
\caption{ {\bf \,\ Variable radio flux densities of \xte.} We list the
  flux densities from all observations of \xte\ with Parkes
  and GBT, as well as detections with the VLA (see also
  ref.~3\protect\nocite{hgb+05}).  The X-ray outburst from \xte\
  occurred between MJDs 52595 and 52662 (ref.~2)\protect\nocite{ims+04}.
  We searched for pulsations from the two pointings nearest to the neutron
  star in the Parkes multibeam pulsar survey\cite{mlc+01} (offset by 8
  and 5\,arcmin, respectively, where the telescope beam has full-width
  at half-maximum of 14.5\,arcmin).  A null result provides the given
  upper limits.  Multiple observations at 1.4 and 1.9\,GHz are listed
  by frequency chronologically in order to aid characterization of flux
  variations.  Following this, we list observations in order of ascending
  frequency.  VLA data are flux-calibrated with high precision using the
  flux density standard 3C286.  Parkes and GBT fluxes are measured using
  the radiometer equation by comparing the area under each average pulse
  profile to the root-mean-square baseline and adopting known observing
  system characteristics.  Absolute uncertainties for these fluxes are
  typically about 30\% for $\nu \la 9$\,GHz, but increase for $\nu \ge
  14$\,GHz (we have applied corrections to account for the opacity of
  the atmosphere).  Owing to poor atmospheric conditions, $S_{42}$ is
  known to within a factor of only about 2. Relative changes in $S_{1.4}$
  can be gauged with approximately 10\% uncertainty, and clear variations
  (by factors of $\approx 2$ on timescales of $\sim 1$\,day) are observed
  at Parkes and also at VLA.  On timescales typically up to one hour there
  is no evidence for intra-day variations in flux from any observation at
  $\nu \la 9$\,GHz.  Conversely, all observations above this frequency
  show evidence for variability on timescales of several minutes.
  For example, on MJD~53828, $S_{8.4}$ was measured in successive 13-min
  scans across a 100-MHz band to be 7.4, 11.2, 9.6, 10.5, 5.1, and
  8.3\,mJy (all uncertainties $\le0.2$\,mJy).  At these high frequencies,
  diffractive interstellar scintillation\cite{ric90,nar92a} is at
  least partly responsible\cite{cl02} for the observed flux variations.
  We denote by superscripts a, b and c the three days on which we obtained
  multi-frequency data, from which the spectral index can be estimated
  with some accuracy: $\alpha \ga -0.5$. }
\label{tab:fluxes}
\end{table*}

\clearpage

\begin{figure*}[t]
\centerline{\psfig{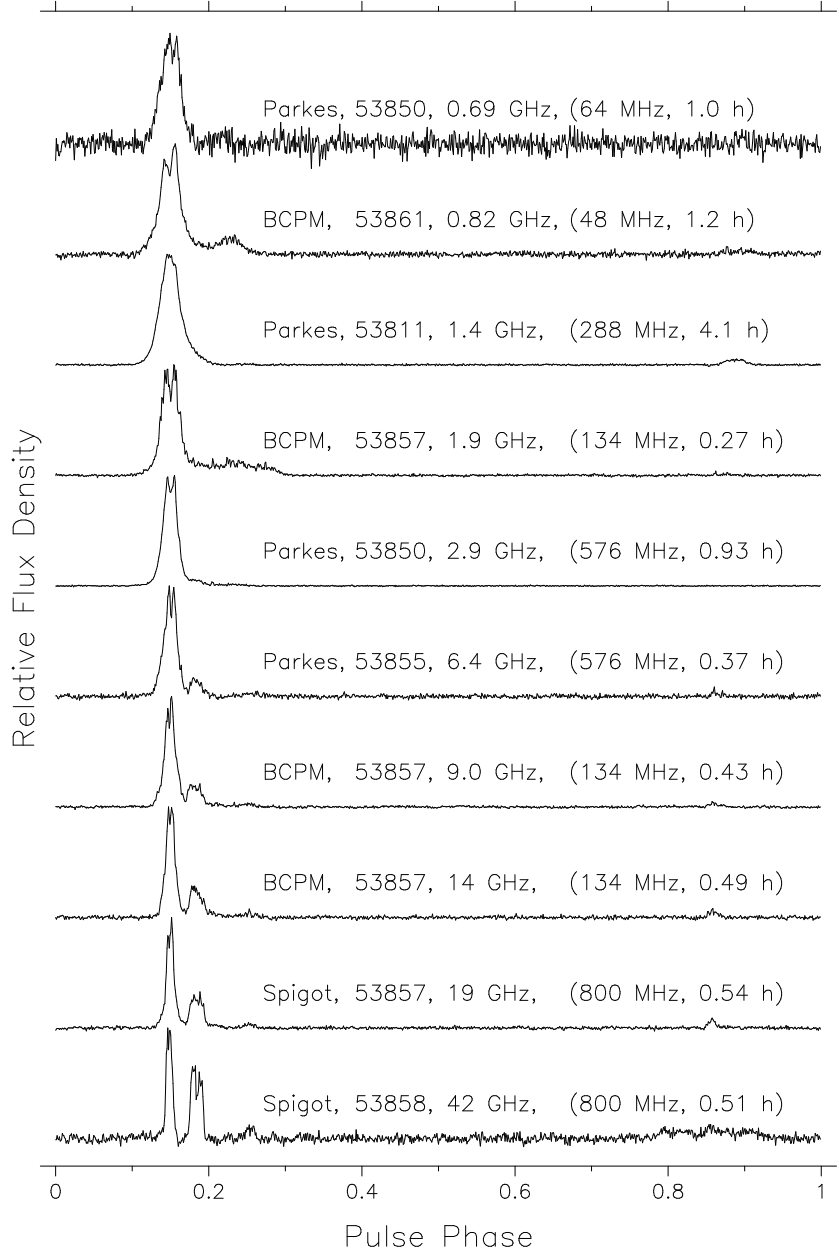}}
\caption{ {\bf \,\ Average radio pulse profiles of \xte\ at frequencies
  0.7--42\,GHz. } Full-period (5.54\,s) profiles are displayed with
  1,024-bin resolution in order of ascending frequency from top to
  bottom, and aligned by fitting the main pulse peak with a gaussian.
  Above each trace we list the equipment, date (MJD), central frequency,
  nominal bandwidth, and integration time used to obtain each profile.
  In some cases a small number of frequency channels corrupted by
  radio-frequency interference have been omitted in constructing the
  profiles.  The Parkes 1.4-GHz (discovery) observation shows a small
  pulse preceding the main pulse by about 0.25 in phase.  This precursor
  pulse is not detected on some days.  These and other changes may be
  related to the well-known emission of separate ``modes''\cite{bac70a}
  displayed by some ordinary pulsars, or could be more exotic, for
  instance due to magnetic field reconfiguration or magnetospheric
  currents varying on timescales of days to weeks.  The separation
  between the two peaks of the main pulse component appears to decrease
  with increasing frequency, possibly akin to what is observed in
  some ordinary pulsars, for which the separation is interpreted as
  reflecting a decreasing emission altitude with increasing frequency.
  For observations at Parkes we used an analogue filterbank with one-bit
  digitization and a high-pass filter of time constant $\approx 0.9$\,s
  that significantly distorts the measured profiles.  We corrected for
  this using the prescription given in ref.~5\protect\nocite{mlc+01}.
  At GBT we used both the BCPM\cite{bdz+97} and Spigot\cite{kel+05},
  and no such coarse artefacts resulted.  }
\label{fig:profs}
\end{figure*}

\clearpage

\begin{figure*}[t]
\centerline{\psfig{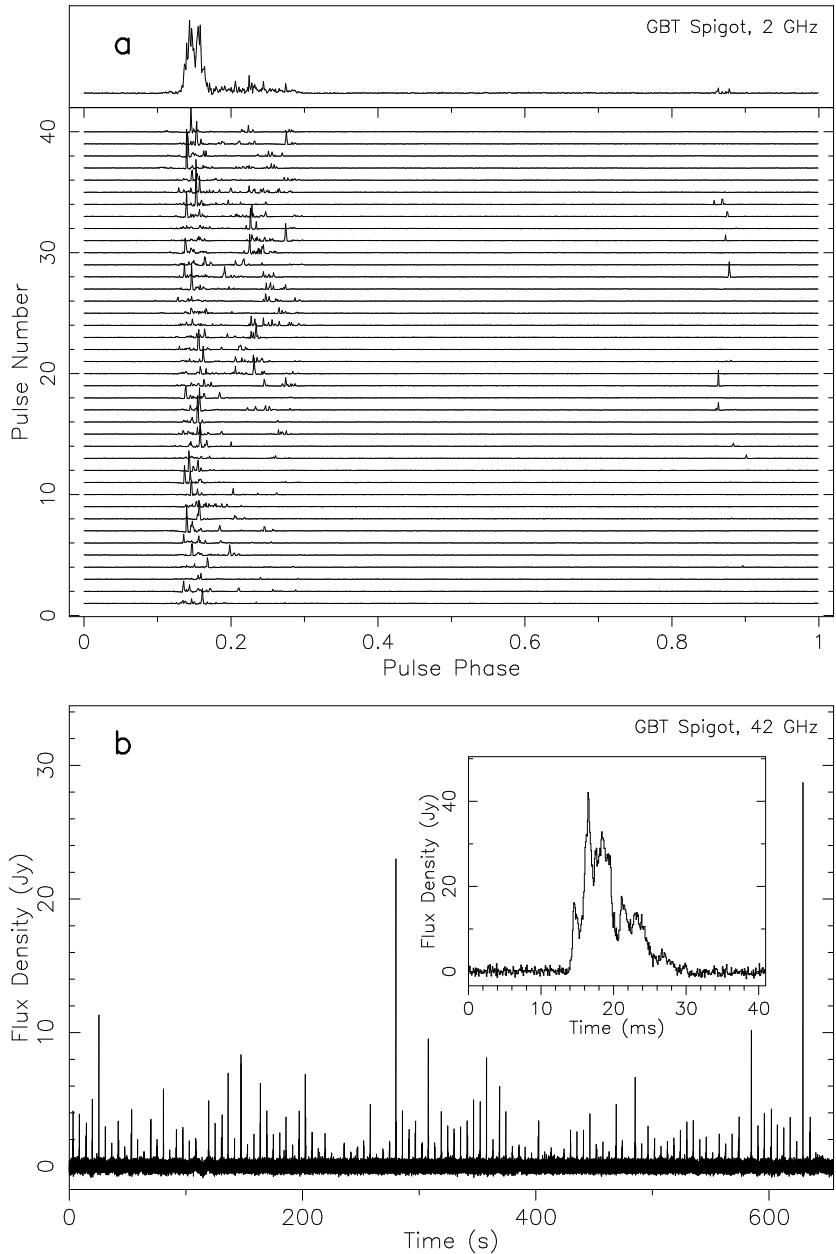}}
\caption{ {\bf \,\ Single pulses from \xte\ at frequencies of 2 and
  42\,GHz. } We detect single pulses from most rotations of the neutron
  star irrespective of frequency.  {\bf a,} We show here a typical set
  of 40 consecutive single pulses from our GBT observation at 2\,GHz on
  MJD~53857, where each row represents the full pulse phase displayed
  with 5.4-ms resolution (1,024 bins).  The sum of all 40 pulses is
  displayed at the top.  Sub-pulses (for which we have found no evidence
  of ``drifts''\cite{dc68}) with typical width $\la 10$\,ms arrive
  at different phases and gradually build up the average profile ---
  which, however, appears different in observations 8 days later, with
  the shoulder at phase 0.2--0.3 essentially missing.  It is unclear
  whether this behaviour is similar in detail to what is observed in
  ordinary pulsars.  {\bf b,} A train of about 115 consecutive single
  pulses detected at a frequency of 42\,GHz with the GBT, displayed
  with 1.3-ms resolution.  For display purposes, we have removed
  large-amplitude power variations with timescales of $\ga 10$\,s,
  probably of atmospheric origin, by high-pass-filtering the data.
  The flux-density scale is uncertain by a factor of about 2.  Inset,
  a 40-ms-long detail of the brightest pulse from the main panel,
  displayed with full 81.92-$\mu$s Spigot resolution.  The pulse has
  structure with features as narrow as $\approx 0.2$\,ms.  }
\label{fig:sp}
\end{figure*}

\end{document}